\pdfoutput=1
\documentclass[aps,prc,reprint,groupedaddress,nofootinbib]{revtex4-1}
\usepackage{amssymb}
\usepackage{amsmath}
\usepackage[colorlinks=true,linkcolor=blue,citecolor=blue, urlcolor=blue]{hyperref}   
\usepackage{tikz}
\usepackage{graphicx}
\usepackage{xspace}
\usepackage[utf8]{inputenc}

\definecolor{uibred}{RGB}{167, 38, 47}
\def\Eq#1{Eq.~(\ref{#1})}

\def\eq#1{(\ref{#1})}

\def\Fig#1{Fig.~\ref{#1}}

\newcommand\pion{\ensuremath{\pi}\xspace}
\newcommand\kaon{\ensuremath{\textrm{K}}\xspace}
\newcommand\pr{\ensuremath{p}\xspace}
\newcommand\cme[1]{\ensuremath{\sqrt{s}\ =\ #1\,\textrm{TeV}}}
\newcommand\cmenn[1]{\ensuremath{\sqrt{s_{\rm NN}}\ =\ #1\,\textrm{TeV}}}
\newcommand\GeVc{\ensuremath{\textrm{GeV}/c}\xspace}

\newcommand\pT{\ensuremath{p_{\textrm{T}}}\xspace}
\newcommand\pt{\ensuremath{p_{\textrm{T}}}\xspace}

\newcommand\mean[1]{\ensuremath{\langle#1\rangle}\xspace}
\newcommand\mbeta{\mean{\beta_{\textrm{T}}}}
\newcommand\td{\ensuremath{d}}
\newcommand\dNps[2]{\ensuremath{\td^{#1}N_{#2}}\xspace}
\newcommand\dEta{\ensuremath{d\eta}\xspace}
\newcommand\dNde{\ensuremath{\dNps{}{\textrm{ch}}/\dEta}\xspace}

\def\p{\mathbf{p}}
\usepackage{tikz}

\def\aprlsection#1{\section{#1}}
\def\prlsection#1{\section{#1}}
\def\App#1{Appendix}
\begin{document}

\title{Temperature and fluid velocity on the freeze-out surface from $\pi$, $K$,  $p$ spectra in pp, p--Pb and Pb--Pb collisions}

\author{Aleksas Mazeliauskas}
\email[]{aleksas.mazeliauskas@cern.ch}
\affiliation{Theoretical Physics Department, CERN, CH-1211 Gen\`eve 23, Switzerland}
\affiliation{Institut f\"ur Theoretische Physik, Universit\"at Heidelberg, Philosophenweg 16, D-69120 Heidelberg, Germany}

\author{Vytautas Vislavicius}
\email[]{vytautas.vislavicius@cern.ch}
\affiliation{Niels Bohr Institutet, Københavns Universitet,
Blegdamsvej 17, DK-2100 Copenhagen, Denmark}

\date{\today}

\begin{abstract}
  We present a new approach to take into account resonance decays in the blast-wave model fits of identified hadron spectra. 
 Thanks to pre-calculated decayed particle spectra, we are able to extract, in a matter of seconds, the
 multiplicity dependence of the single freeze-out temperature $T_{\rm fo}$, average fluid velocity $\left<\beta_{\rm T}\right>$,
 velocity exponent $n$, and the volume $dV/dy$ of an expanding fireball.
In contrast to blast-wave fits without resonance feed-down,
our approach results in a freeze-out
temperature of $T_{\text{fo}}\approx 150\,\text{MeV}$, which has only weak
dependence on multiplicity and collision system. Finally, we discuss separate chemical and kinetic freeze-outs separated by partial chemical equilibrium.
\end{abstract}

\maketitle
\prlsection{Introduction}The relativistic hadron collisions explore the properties
of dense nuclear matter at temperatures
several times higher than that of the pseudo-critical QCD temperature  $T_c=
156.5\pm 1.5\,\text{MeV}$~\cite{Bazavov:2018mes}, i.e.\ the state of deconfined
quarks and gluons.
Remarkably, the study of produced hadron and light nuclei abundances indicates
an apparent thermal particle production at constant chemical freeze-out temperature $T_\text{chem}\approx T_c$,
as shown by fits of statistical hadronization model (SHM)~\cite{Andronic:2017pug}.
Furthermore, phenomenological models based on viscous fluid description of the quark-gluon plasma (QGP) expansion
successfully reproduce
many soft hadronic observables%
~\cite{Heinz:2013th,Teaney:2009qa,Luzum:2013yya,Gale:2013da,deSouza:2015ena}.
Global fits to experimental data can then be used to extract
the model parameters and the transport properties of dense QCD matter~\cite{Bernhard:2016tnd}.

One of the earliest and simplest models of hadron production from a flowing medium is the blast-wave model~\cite{Schnedermann:1993ws}.
It is based on calculating particle emission from a parametrized freeze-out surface of temperature $T_\text{fo}$ and radial velocity profile $\beta_T(r)$.
The primary particle spectra is taken to be thermal in the local rest-frame of the fluid.
Then the experimentally observed hadrons, e.g.\ pions, kaons or protons,
are calculated by adding the decay feed-down from the short-lived primary resonances to the initial
thermal abundances.
In general, freeze-out with only direct decays gives
a reasonably good description of the data~\cite{Broniowski:2001we,Retiere:2003kf, Ryu:2017qzn}, but
neglects possible re-scattering and re-generation of hadrons,
which can be modeled by hadronic after-burners~\cite{Bass:1998ca,Petersen:2018jag}.
The blast-wave model can be simplified even further by using thermal spectra of
pions, kaons and protons
to directly fit the measured particle spectra.
As decay feed-down significantly
modifies the magnitude and  momentum dependence of
distributions, individual normalizations are introduced for each particle
species and the momentum range for the fit is restricted~\cite{Abelev:2013vea}.
In this case the extracted freeze-out temperature and radial velocity profiles are interpreted
as temperature and velocity at the kinetic particle freeze-out. This is the routine
analysis procedure performed for measured identified particle spectra as a convenient
way to characterize and compare the soft particle production at different centralities
and collision systems~\cite{Abelev:2008ab,Abelev:2013vea, Chatrchyan:2013eya, Abelev:2013haa, ALICE:2017jyt}.

In this paper we present a new procedure for the blast-wave model fits, which
includes the feed-down from resonance
decays. We are certainly not the first to include resonance decays in the blast-wave model,
as it was done already in~\cite{Schnedermann:1993ws} and other studies~\cite{Broniowski:2001we, Begun:2013nga, Melo:2015wpa, Melo:2019mpn}.
 However, up to now the generation of primary thermal hadrons and their decays were two separate
 steps, the latter computed by either Monte-Carlo generators~\cite{Torrieri:2004zz,Amelin:2006qe,Tomasik:2008fq,Chojnacki:2011hb},
or semi-analytic treatments of decay integrals~\cite{Sollfrank:1991xm,Sollfrank:1990qz}.
This amounts to considerable computational costs, as
for each set of model parameters a large number  of primary
hadron spectra need to be generated and then decayed through thousands of decay channels~\cite{Tanabashi:2018oca}.
 Instead we use recently published method of efficient calculation of direct resonance
 decays~\cite{Mazeliauskas:2018irt, FastReso}. The technique is based on first calculating the resonance decays in
 fluid rest-frame and only then finding the final particle
 spectra for a fluid cell moving with some velocity $u^{\mu}(x)$.
 In this approach  the primary resonances and decays need to be evaluated only once for a given temperature and chemical potential, which greatly simplifies and speeds up the fitting procedure.

\clearpage
\prlsection{Blast-wave fit with resonance decay feed-down\label{sec:model}}In a boost invariant blast-wave freeze-out model, particles are produced from
a constant time $\tau_\text{fo}$
 and temperature $T_\text{fo}$ freeze-out surface with
transverse radius $R$ and a power-like velocity profile~\cite{Schnedermann:1993ws}
\begin{equation}
\beta_T\equiv \frac{u^r}{u^\tau} = \beta_\text{max} \left( \frac{r}{R} \right)^n
.\end{equation}
Thermal particle production from a fluid cell of temperature $T_{\text{fo}}$ moving with a 4-velocity $u^\mu$ can be
calculated according to the Cooper\nobreakdash-Frye formula~\cite{Cooper:1974mv},
\begin{equation}
E_\p\frac{d N}{d^3 \p} = \frac{\nu}{(2\pi)^3}\int_\sigma f\left(-u^\nu p_\nu, T_{\text{fo}}, \mu\right)
p^\mu d\sigma_\mu.\label{eq:CF}
\end{equation}
Here $\nu=(2S+1)$ is the spin degeneracy, $d\sigma_\mu$ is the freeze-out surface element (for blast-wave surface  $d\sigma_\mu=(\tau_\text{fo}d\eta r d\phi dr,0,0,0)$, $\mu$ is the chemical potential, $\bar E_p = - u^\nu p_\nu$ is the
fluid-frame energy of the particle, and $f$ is the thermal particle distribution function.

The unstable resonances produced on the freeze-out surface according to \Eq{eq:CF} decay
and non-trivially modify the momentum spectra of long lived-hadrons. It was shown
in Ref.~\cite{Mazeliauskas:2018irt} that decay feed-down modification for thermally produced hadrons
can be captured by two scalar distribution functions, $f^\text{eq}_{1}$ and $f^\text{eq}_{2}$~\footnote{
  $f_{i=1,2}^{\text{eq}}$ are components of irreducible de-composition under rotations of the decayed particle spectra  in the fluid rest-frame~\cite{Mazeliauskas:2018irt}},
which generalize the Cooper-Frye freeze-out integral to
\begin{align}
E_\p\frac{d N}{d^3 \p} &= \frac{\nu}{(2\pi)^3}\int_\sigma\left[ f^{\text{eq}}_{1}\left(p^\mu-\bar E_\p u^\mu\right) + f^\text{eq}_{2} \bar E_\p u^\mu \right] d\sigma_\mu.
\end{align}
Given a list of resonances and decay channels functions $f^\text{eq}_{i=1,2}\left(-u^\nu p_\nu, T_{\text{fo}}, \mu\right)$
can be easily computed using publicly available code~\cite{FastReso}.
For the azimuthally symmetric and boost-invariant blast-wave surface the decayed particle
spectra simplifies to a 1-dimensional integral~\cite{Mazeliauskas:2018irt}
\begin{align}
&\frac{d N}{2\pi p_Tdp_Tdy} = \frac{\nu}{(2\pi)^3} \int_0^R\! dr \; \tau_\text{fo}r\, K^\text{eq}_1\left(p_T, \beta_T(r) \right). \label{eq:BW}
\end{align}
The freeze-out kernel $K_1^\text{eq}(p_T, \beta_T, T_\text{fo},\mu)$ can be evaluated in advance
for a range of values $(p_T, \beta_T)$ by azimuthal and space-time rapidity integrals of functions $f_{i=1,2}^\text{eq}$
\begin{align}
K^\text{eq}_1(p_T,\beta_T)  =  & \int_0^{2\pi} \!d\phi \int_{-\infty}^\infty\! d\eta  \left\{ f^\text{eq}_1(\bar E_\p) m_T \cosh(\eta)
\right.\nonumber\\
&
\left.
+ \left(f^\text{eq}_2(\bar E_\p) - f^\text{eq}_1(\bar E_\p)\right) \bar E_\p u^\tau \right\},
\end{align}
where $\bar E_\p = m_T u^\tau \cosh(\eta) - p_T u^r \cos \phi$ is the particle energy in the fluid rest-frame, the transverse mass is defined as $m_T=\sqrt{p_T^2+m^2}$, and radial 4-velocity is $u^r = \beta_T/\sqrt{1-\beta_T^2} $.
Equation \eq{eq:BW} should be compared with the analogous equation in the
standard blast-wave fit, where the thermal freeze-out kernel is given by the corresponding integral of
 the Boltzmann
distribution~\cite{Schnedermann:1993ws}
\begin{align}
K^\text{th}_1(p_T, \beta_T)& = 4\pi m_T \mathcal I_0\left(\frac{p_T u^r}{T_\text{fo}}\right)\mathcal K_1\left(\frac{m_T u^\tau}{T_\text{fo}}\right).
\end{align}
The crucial difference is that our freeze-out kernel  $K^\text{eq}_1(p_T,
\beta_T)$ already contains the feed-down contributions from the unstable
resonances. Therefore different particles produced from the same freeze-out
surface have the same normalization in \Eq{eq:BW}, namely the freeze-out volume
per rapidity $dV/dy = \tau_\text{fo} \pi R^2$ (in the lab-frame).

Finally we note that although in \Eq{eq:BW} we considered a very specific freeze-out surface, the procedure can be straightforwardly
extended to more complicated freeze-out surfaces by introducing additional freeze-out kernels~\cite{Mazeliauskas:2018irt}.

\prlsection{Partial chemical equilibrium}%
To allow for separate chemical and kinetic freeze-outs we employ the partial
chemical equilibrium (PCE) model~\cite{Bebie:1991ij,Hirano:2002ds}. In this model the quasi-stable hadrons, $b$,
maintain an approximate kinetic equilibrium through elastic scatterings, but
the particle ratios are fixed at the chemical freeze-out temperature $T_\text{chem}$.
Then at the kinetic freeze-out the distribution function of resonance $a$ is given by a thermal
distribution at temperature $T_\text{fo}=T_\text{kin}$, but with
chemical potential $\tilde \mu_a(\mu_b) = \sum_b N_{a\rightarrow b}\mu_b$, where $N_{a\rightarrow b}$
is the number of decay products $b$, and  $\mu_b$ is the chemical potential of the
quasi-stable species $b$. Assuming ideal hydrodynamic evolution between the chemical and kinetic freeze-outs, chemical potentials  $\mu_b$
are such that entropy
per quasi-stable particle $b$ is conserved, i.e.\ we need to solve the implicit equation
for $\mu_b$
 \begin{equation}
   \frac{\sum_a N_{a\rightarrow b} n_a(T_\text{chem},0)}{\sum_a s_a(T_\text{chem},0)} = \frac{\sum_a N_{a\rightarrow b} n_a(T_\text{kin},\tilde \mu_a(\mu_b))}{\sum_a s_a(T_\text{kin}, \tilde \mu_a(\mu_b))}
,\end{equation}
where the sum goes over all resonance species $a$ and $n_a,s_a$ are the number and entropy density
for an ideal gas of resonance species  $a$.
Then the freeze-out kernel for partial chemical equilibrium
\begin{equation}
  K^\text{eq}_1(p_T,\beta_T; T_\text{kin}, \mu(T_\text{kin}, T_\text{chem}))\label{eq:KPCE}
.\end{equation}
can be computed by decaying hadrons at the kinetic freeze-out $(T_\text{kin}, \tilde \mu_a(\mu_b))$ using the same techniques.~\cite{Mazeliauskas:2018irt}.

\prlsection{Setup\label{sec:setup}}We evaluated the irreducible scalar
distributions $f^{\text{eq}}_{1,2}$ for $\pi$, $K$, $p$, $\Lambda$, $\Xi$ and
$\Omega$ using the publicly available code \texttt{FastReso}~\cite{FastReso}.
We use a recent list of resonances and decay channels (including less 
established states)~\cite{Alba:2017mqu, Alba:2017hhe, parotto_private}
derived from 2016 edition of Particle Data Group book~\cite{Patrignani:2016xqp}.
In total, we  perform $3291$ 2-body and  $513$ 3-body strong and electromagnetic
decays for 739 resonances with masses up to $3.0\, \text{GeV}/c^2$.
For the single
freeze-out fits of $\pi,K,p$ spectra, we evaluated the freeze-out kernels $K^{\text{eq}}_1$ for temperatures in $[130-180]$ MeV interval in 0.5 MeV steps and kept baryon chemical potential $\mu_B=0$. For calculating PCE
freeze-out kernels, \Eq{eq:KPCE}, we followed Ref.~\cite{Huovinen:2007xh}
and conserved the particle number of $\pi$, $K$, $\eta$, $\omega$, $p$, $n$, $\eta'$, $\phi$, $\Lambda$, $\Sigma$, $\Xi$, $\Lambda(1520)$, $\Xi(1530)$, and $\Omega
$ hadrons. Then we fixed the chemical freeze-out temperature and varied $T_{\text{kin}}$ in 1 MeV steps in $[100-200]$ MeV interval. 

\prlsection{Results\label{sec:results}}The data considered in this work has been published by the ALICE Collaboration~\cite{Aamodt:2008zz}
and includes \pion, \kaon and \pr spectra measured as a function of
centrality and multiplicity in pp collisions at \cme{7}~\cite{Acharya:2018orn}, p--Pb
collisions at \cmenn{5.02}~\cite{Adam:2016dau} and Pb--Pb collisions at \cmenn{2.76}~\cite{Abelev:2013vea}.
The blast-wave model parameters are extracted by simultaneously
fitting the $\pi,K,p$ spectra in the transverse momentum range $0.5 < \pT <
3.0\,(\GeVc)$, where each data point is considered with a weight given by the
statistical and systematic uncertainties summed in quadrature.
We checked that extracted parameters are insensitive to the choice of momentum range
for central and mid-central Pb--Pb collisions and p--Pb and pp results show only
modest dependence on the fit ranges (see \App{sec:app}).
We find a very good fit for Pb--Pb spectra with $\chi^{2}/\text{dof}$ in the range $0.5-2.9$.
For smaller collision systems $\chi^{2}/{\rm dof}$ grows from 1.5 in the highest multiplicity p--Pb collisions
to 8.2 in the lowest multiplicity pp bin. For completeness the tables of the best fit values, their uncertainties
and $\chi^{2}/{\rm dof}$ are given in \App{sec:app}.

\begin{figure}
  \includegraphics[width=1.0\columnwidth]{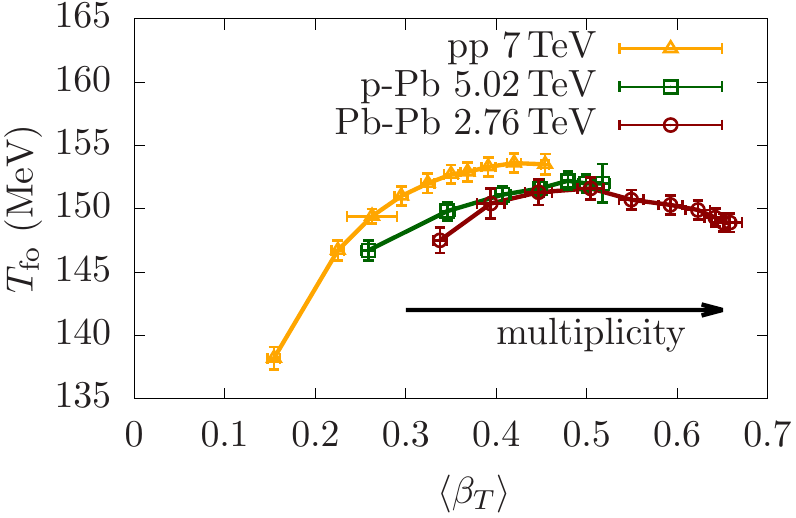}
\caption{\label{fig:plottkinvsbetat}
  Blast-wave freeze-out temperature $T_\text{fo}$ versus mean transverse
  velocity $\left<\beta_T \right>$ in pp, p--Pb and Pb--Pb collisions including the feed-down of resonance decays. Spectra of $\pi$,
  $K$ and $p$ are fitted  in  $0.5< p_T<3.0\,(\text{GeV}/c)$ momentum
  range. Error bars correspond to fit parameter uncertainties.} \end{figure}

In \Fig{fig:plottkinvsbetat} we show the extracted freeze-out temperature $T_\text{fo}$
plotted as a function of mean radial velocity $\left<\beta_T\right> \equiv 2 \beta_\text{max}/\left( 2+n \right) $ and
collision system. All systems show stronger radial flow with increasing multiplicity, but only modest temperature dependence. We see that for Pb--Pb collisions the freeze-out temperature is in the
$148\text{--}152\,{\rm MeV}$ range~\footnote{We note that including additional resonances states from PDG2016+ systematically lowers the extracted freeze-out temperature, but the radial flow remains unchanged.} and close to the chemical freeze-out
temperature $T_\text{chem}= 156.5\pm 1.5\,\text{MeV}$ obtained by the
statistical model fitted to light and multi-strange hadrons in the most central
collisions~\cite{Andronic:2017pug}. Studies of centrality dependence of
thermal model parameters also find weak temperature dependence, but the temperature is higher than
extracted in out fits~\cite{Becattini:2014hla, Vovchenko:2019kes, Vovchenko:2019pjl}. We note that in our model we use only
$\pi,K,p$ spectra and do not introduce baryon chemical potential, canonical suppression or strangeness undersaturation effects.
In smaller systems we observe similar values of freeze-out temperature,
and in the case of p--Pb collisions we find overlapping freeze-out temperature and radial flow values compared to Pb--Pb.
Finally, pp collisions tend to have smaller average radial flow, but the temperature dependence is comparable to other systems, except for the lowest multiplicity bin.

Our results in \Fig{fig:plottkinvsbetat} are noticeably different from
the usual blast-wave fits without resonance decays,
which show strong
freeze-out temperature dependence on centrality in Pb--Pb collisions~\cite{Abelev:2013vea}.
To understand this difference, we repeated the fits for Pb--Pb data by allowing independent normalizations
of each particle spectrum.
We found that $\chi^{2}/\text{dof}$ of such a fit
does not have a well localized
minimum in the freeze-out temperature $T_{\text{fo}}$, but instead has a
very shallow region over the entire considered range of temperature. 
In fact, it was already shown in Ref.~\cite{Schnedermann:1993ws}, that
due to the feed-down of heavier resonances, pions respond to radial flow much
like heavy particles and equally good fits to particle spectra can be  obtained
for different values of freeze-out temperature. 
In contrast, the blast-wave fit without resonance decays, erroneously singles out a particular combination
of temperature and radial flow. 

\begin{figure}
  \centering
\includegraphics[width=1.0\columnwidth]{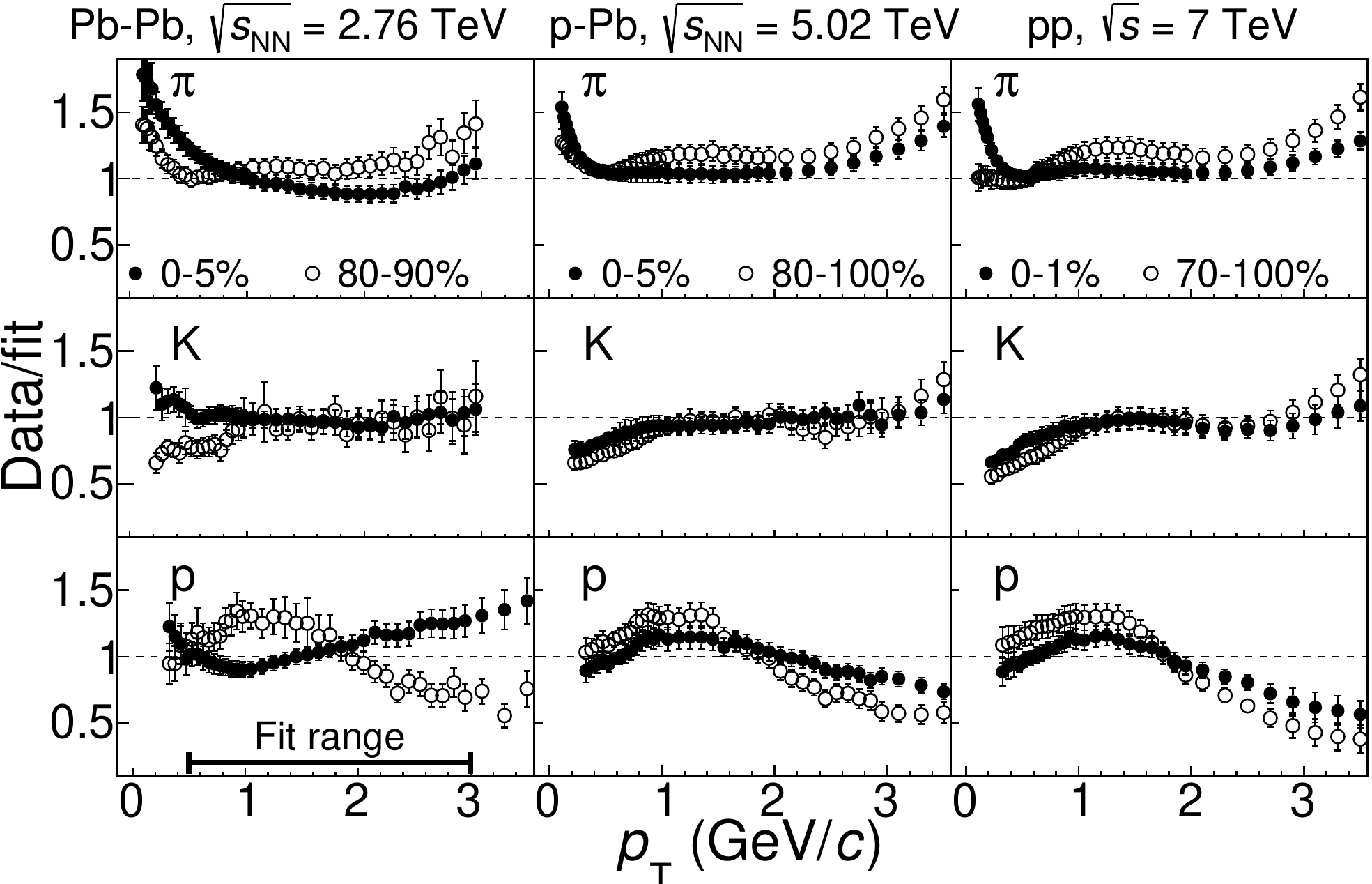}
  \caption{Transverse momentum spectra of $\pi, K,p$ divided by the blast-wave fit with resonance decays at most
  central (full symbols) and peripheral (open symbols) centrality classes. Error bars correspond to
statistical and systematic uncertainties of the data summed in quadrature.}
  \label{fig:Figures-DataToFitRatios}
\end{figure}
\begin{figure}
\centering
\includegraphics[width=1.0\columnwidth]{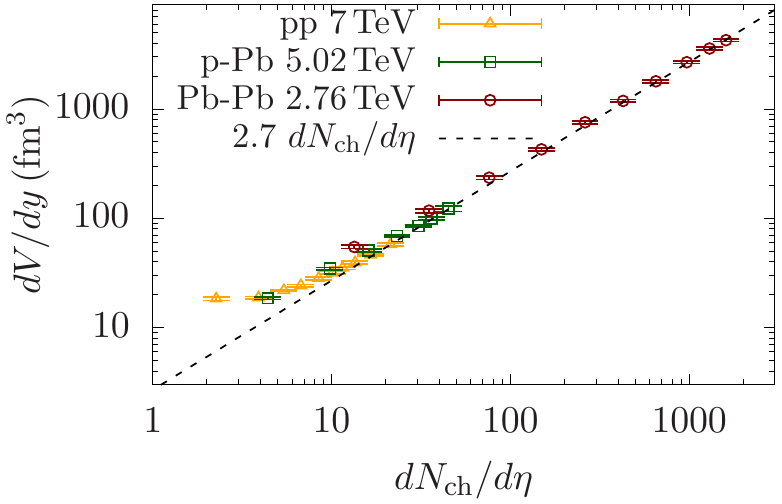}
\caption{Extracted freeze-out volume  per rapidity $dV/dy$ (in the lab-frame) as a function of multiplicity for different collision systems. Error bars correspond to fit parameter uncertainties.}
\label{fig:plotTnbeta}
\end{figure}
Next we study the ratios of measured hadron spectra to the improved blast-wave model fits, which  are shown in
\Fig{fig:Figures-DataToFitRatios} for different collision systems and most central and most peripheral centrality bins.
We find that \pion, \kaon and $p$ spectra are described by the model
within $\approx 2\text{--}4\, \sigma$ range for momenta $0.5<\pT<3.0\,({\rm GeV}/c)$,
suggesting that the primary hadrons are emitted from a fluid expanding with common velocity field.
The ratios are flat for pions and kaons, but the proton
spectra-to-fit ratio shows a residual evolution with \pT, which can be attributed to the
rescatterings in the hadronic gas phase, which
generally boost heavier particles towards higher momenta~\cite{Ryu:2017qzn}.

We would like to emphasize that including resonance decays in the blast-wave fits
allows us to use a single
normalization factor for all particle species. The extracted factor can be interpreted
as the freeze-out volume per unit rapidity of the fireball and is
proportional to the overall multiplicity \dNde, as shown in
\Fig{fig:plotTnbeta}.
\begin{figure}
  \centering
    \begin{tikzpicture}
    \node (image) at (0,0) { \includegraphics[width=1.0\columnwidth]{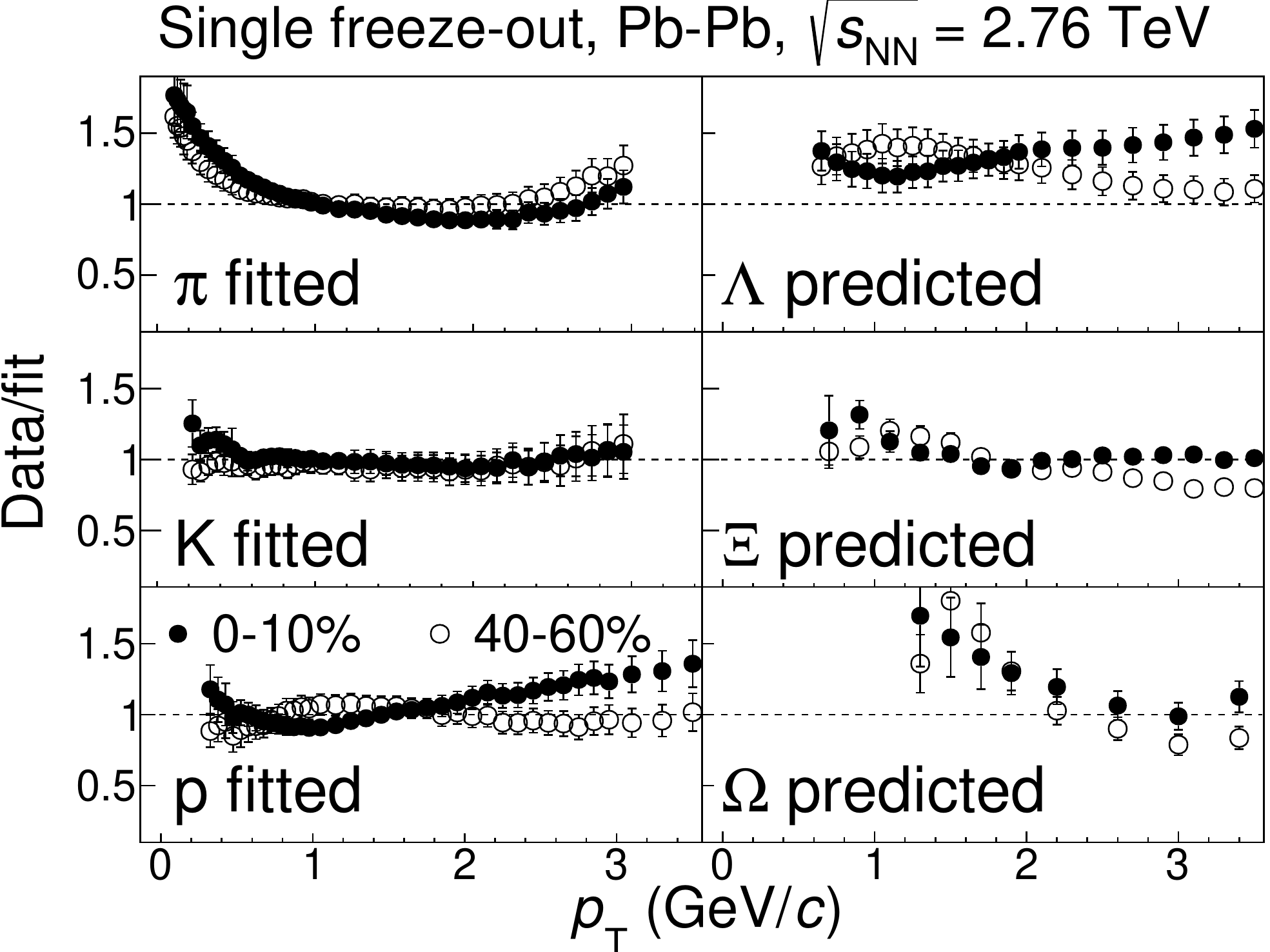} };
    \end{tikzpicture}
    \caption{(left) Data to fit ratios of $\pi,K,p$  spectra in single freeze-out blast-wave model. (right) Data to prediction ratios of $\Lambda,\,\Xi,\,\Omega$ spectra obtained using the same blast-wave fit parameters.
 Error bars correspond to
statistical and systematic uncertainties of the data summed in quadrature.
}
  \label{fig:Figures-DataToFitRatios_MS}
\end{figure}
Because of fixed normalization, the  extracted blast-wave model parameters $\beta_\text{max}$,
$T_{\rm fo}$, $n$ and $dV /dy$ from fits to \pion, \kaon and \pr spectra can be used
to predict heavier hadrons, such as $\Lambda,\,\Xi,\,\Omega$~\cite{ALICE:2017jyt,Abelev:2013haa,Adam:2015vsf,Abelev:2013xaa,ABELEV:2013zaa}.
We show the data to model ratios in \Fig{fig:Figures-DataToFitRatios_MS} ($\pi,K,p$
spectra were refitted in the same centrality bins).
The model predictions for $\Xi$ and $\Omega$ are good and only
$\Lambda$ spectra is somewhat under-predicted (similar discrepancies are also seen in full hydrodynamic simulations \cite{Ryu:2017qzn}).

\begin{figure}
  \centering
    \begin{tikzpicture}
    \node (image) at (0,0) { \includegraphics[width=1.0\columnwidth]{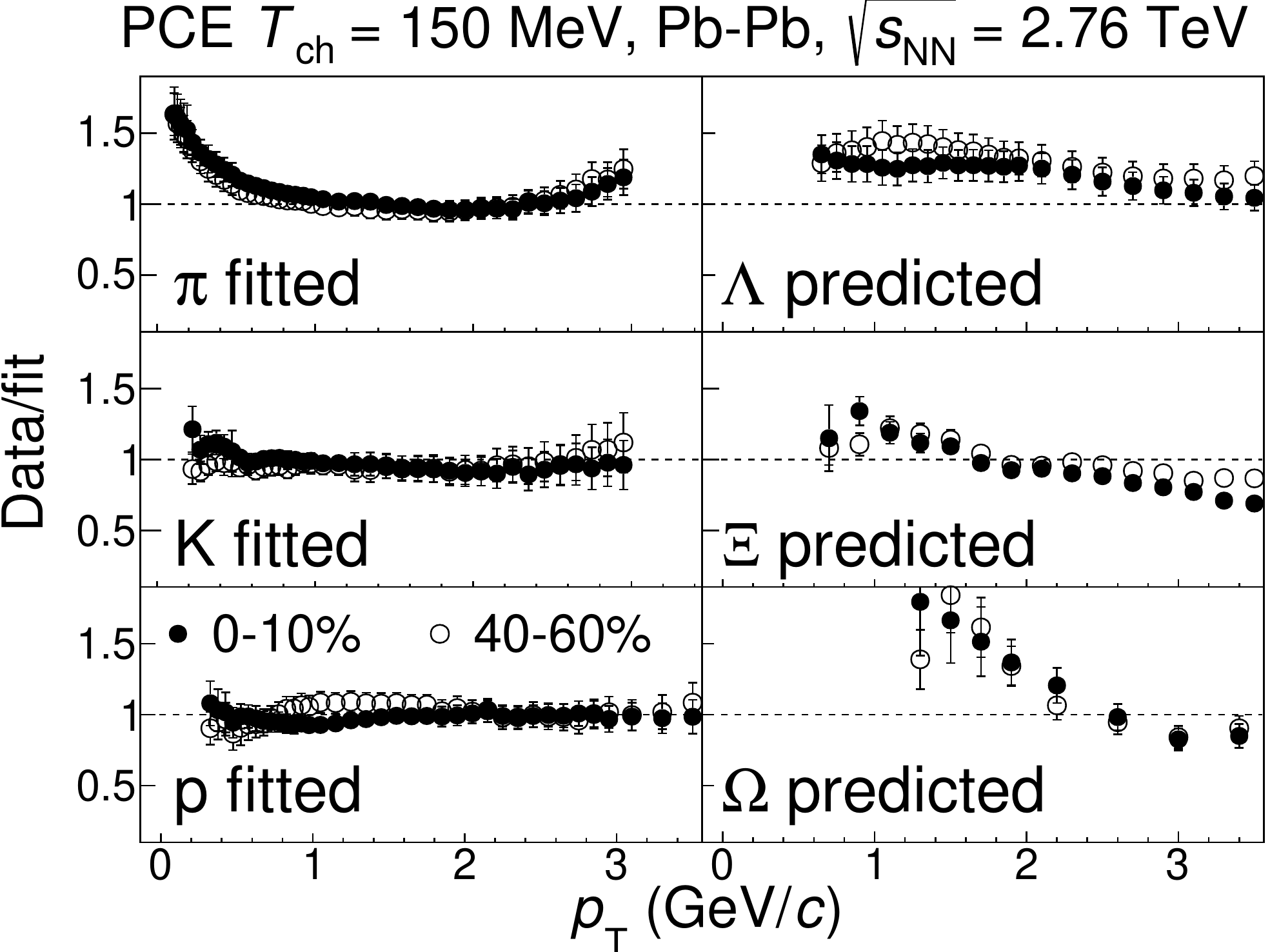} };
    \end{tikzpicture}
    \caption{(left) Data to fit ratios of $\pi, K,p$ spectra in partial chemical equilibrium model with chemical freeze-out at $T_{\text{chem}}=150\,\text{MeV}$. (right) Data to prediction ratios of $\Lambda,\,\Xi,\,\Omega$ obtained using the same blast-wave fit parameters.
 Error bars correspond to
statistical and systematic uncertainties  of the data summed in quadrature.
}
  \label{fig:Figures-DataToFitRatios_MSPCE155}
\end{figure}

Finally, we consider the blast-wave fits with distinct chemical and kinetic freeze-outs using partial chemical equilibrium model.
We fix the chemical freeze-out temperature  $T_{\text{chem}}$ (and hence particle ratios)
and vary the kinetic freeze-out temperature $T_{\text{kin}}$. In \Fig{fig:Figures-DataToFitRatios_MSPCE155}.
we show the spectral ratios for Pb--Pb collisions at two different centralities for $T_\text{chem}=150\,\text{MeV}$. The data to model
ratio for protons becomes flat in 0-10\% centrality with the kinetic freeze-out temperature $T_\text{kin}=126\pm2\, \text{MeV}$,
while for 40-60\% centrality the kinetic and chemical freeze-out temperatures become approximately equal.
For pp and p--Pb collisions, we  do not obtain a convergent fit, with $T_\text{kin}>T_\text{chem}$ reaching the
upper limit ($200\,\text{MeV}$) of the fitting range. This points out to the fact that
additional observables, e.g.\ short lived resonances, are needed to differentiate
 chemical and kinetic freeze-outs~\cite{Vovchenko:2019aoz}.

\prlsection{Conclusions and Outlook\label{sec:summary}}We performed the multiplicity
and collision system analysis of identified hadron spectra using the blast-wave model with resonance decays
for pp, p--Pb and Pb--Pb collisions at the LHC. Thanks to the inclusion
of decay feed-down, we were able to fit $\pi, K,p$ spectra in a wide momentum range $0.5<p_T < 3.0\, (\GeVc)$
and extract the common freeze-out volume, freeze-out temperature and the radial flow parameters.
In contrast to traditional blast-wave fits,
our fits take into account both the shape and relative normalization of particle spectra. Consequently, our method
produces a single freeze-out temperature, which is relatively insensitive to the multiplicity, system size or fitting ranges.
By using independent normalization of spectra we checked that with decay feed-down, only the shape of pion, kaon and proton spectra alone is not sufficient to unambiguously determine the freeze-out temperature and radial flow~\cite{Schnedermann:1993ws}.
Our fit of $\pi,K,p$ spectra is in $\approx 2\text{--}4\,\sigma$ agreement with experimental data
in the fitting range  $0.5<p_T < 3.0\, (\GeVc)$ for
 all multiplicity classes and  collision systems. Furthermore, using the extracted
 freeze-out volume we were able to predict multi-strange particle  spectra ($\Lambda,\Xi, \Omega$) without additional model parameters.
Finally, introducing separate kinetic and chemical freeze-outs separated by partial chemical
equilibrium phase slightly improved
the fit of proton spectra in central 0-10\% Pb--Pb collisions with $T_\text{kin}\approx 126\pm 2 \,\text{MeV}$, which grows towards and past $T_\text{chem}$ in more peripheral collisions.
However the procedure
does not result in
physical parameter values in small multiplicity collisions. We conclude that particle spectra beyond the long lived $\pi, K, p$ are needed to constrain the kinetic freeze-out temperature.

The most significant aspect of our study is the practical demonstration that simple
data analysis including the important effect of decay feed-down can be done in a computationally efficient
way. By first calculating the decayed particle spectra in the
fluid frame for a range of freeze-out parameters~\cite{Mazeliauskas:2018irt,FastReso},
 we were able to perform the blast-wave fit analysis of particle spectra in a matter of seconds.
This practical approach opens up a way for simple, but realistic studies of particle
production in hadronic collisions using parametrized freeze-out surfaces. Useful
physical insight could be gained by studying the shape of freeze-out surface in small collision systems~\cite{Heinz:2019dbd},
the effect of viscous corrections to particle spectra and elliptic
flow~\cite{Dusling:2009df,Teaney:2013gca,Ryu:2017qzn,Huovinen:2001cy}, the
freeze-out criteria~\cite{BraunMunzinger:2003zz,Ramamurti:2018urm} and inclusion of additional observables~\cite{Florkowski:2019voj,Andronic:2019wva, Bellwied:2018tkc, Bluhm:2018aei,Motornenko:2019jha}.
These studies would clearly complement the on-going multi-parameter hydrodynamic modeling of heavy-ion
collisions, which ultimately can be used to determine the properties of the QGP~\cite{Bernhard:2016tnd,Devetak:2019lsk}.

\begin{acknowledgments}
      A.M.\ thanks D.~Devetak, A.~Dubla, S.~Floerchinger, E.~Grossi, S.~Masciocchi, I.~Selyuzhenkov, and D.~Teaney
 for the discussions and collaboration on related projects.
The authors thank P.~Parotto for sharing their PDG2016+ resonance and decay lists.
The authors also thank
 A.~Andronic, P.~Braun-Munzinger, U.~Heinz, P.~Huovinen, W.~Florkowski, A.~Kalweit, J.~Stachel, K.~Reygers,
 and V.~Vovchenko for useful discussions and comments.
This work was supported in part by the German Research Foundation (DFG) Collaborative Research Centre “SFB 1225
(ISOQUANT)” (A.M.), the Danish Council for Independent Research---Natural Sciences,
the Carlsberg Foundation, and the Danish National Research Foundation (DNRF) (V.V.).
V.V.\ thanks the Institute for Theoretical Physics at Heidelberg University for the hospitality during a
short term visit.
\end{acknowledgments}

\bibliography{master.bib}

\appendix
\aprlsection{Best fit parameters and fit ranges\label{sec:app}}In Tables \ref{tab:pp}, \ref{tab:pPb} and \ref{tab:PbPb}
we summarize the best fit values and uncertainties
for different collision systems and centralities for single freeze-out blast-wave fits. In addition in Tables~\ref{tab:PbPb276TeV_MS} and  \ref{tab:PbPb276TeV_MS_PCE150} we report the fit parameters for Pb--Pb collisions in different centrality bins within single freeze-out and partial chemical equilibrium models.

\begin{table}[h]
\caption{\label{tab:pp} Results for $\pi$, $K$, $p$ combined blast-wave fit with resonance decays  for pp \cme{7} data  in momentum range $0.5<p_{ T} <3.0\,(\text{GeV}/c)$.}
\begin{tabular*}{\linewidth}{l@{\extracolsep{\fill}}c@{\extracolsep{\fill}}c@{\extracolsep{\fill}}c@{\extracolsep{\fill}}c@{\extracolsep{\fill}}c}
\hline
\hline
Centrality & $\left<\beta_T\right>$  & $T_\text{fo}$ (MeV) & $n$ & $dV/dy$ (fm${}^3$) & $\chi^2/\text{dof}$ \\
\hline
0-1\% & 0.454$\pm$0.004 &  154$\pm$1 & 2.00$\pm$0.04 &57.7$\pm$1.7   &2.2\\
1-5\% & 0.420$\pm$0.008 &  154$\pm$1 & 2.32$\pm$0.08 &46.1$\pm$1.3   &3.1\\
5-10\% & 0.391$\pm$0.008 &  153$\pm$1 & 2.62$\pm$0.09 &39.3$\pm$1.1   &3.8\\
10-15\% & 0.368$\pm$0.008 &  153$\pm$1 & 2.91$\pm$0.10 &35.0$\pm$1.0   &4.4\\
15-20\% & 0.350$\pm$0.008 &  153$\pm$1 & 3.16$\pm$0.11 &31.7$\pm$0.9   &4.9\\
20-30\% & 0.324$\pm$0.007 &  152$\pm$1 & 3.56$\pm$0.12 &28.2$\pm$0.8   &5.7\\
30-40\% & 0.295$\pm$0.007 &  151$\pm$1 & 4.10$\pm$0.15 &24.2$\pm$0.6   &6.5\\
40-50\% & 0.263$\pm$0.028 &  149$\pm$1 & 4.85$\pm$0.71 &21.8$\pm$0.3   &7.6\\
50-70\% & 0.225$\pm$0.007 &  147$\pm$1 & 5.95$\pm$0.25 &18.7$\pm$0.5   &8.0\\
70-100\% & 0.154$\pm$0.007 &  138$\pm$1 & 9.43$\pm$0.50 &18.3$\pm$0.6   &8.2\\
\end{tabular*}
\end{table}

\begin{table}[h]
\caption{\label{tab:pPb} Results for $\pi$, $K$, $p$ combined blast-wave fit with resonance decays  for p--Pb \cmenn{5.02} data in momentum range $0.5<p_{ T} <3.0\,(\text{GeV}/c)$.}
\begin{tabular*}{\linewidth}{l@{\extracolsep{\fill}}c@{\extracolsep{\fill}}c@{\extracolsep{\fill}}c@{\extracolsep{\fill}}c@{\extracolsep{\fill}}c}
\hline
\hline
Centrality & $\left<\beta_T\right>$  & $T_\text{fo}$ (MeV) & $n$ & $dV/dy$ (fm${}^3$) & $\chi^2/\text{dof}$ \\
\hline
0-5\% & 0.52$\pm$0.01 &  152$\pm$2 & 1.44$\pm$0.05 & 122$\pm$7   &1.5\\
5-10\% & 0.50$\pm$0.01 &  152$\pm$1 & 1.57$\pm$0.06 & 100$\pm$3   &2.0\\
10-20\% & 0.48$\pm$0.01 &  152$\pm$1 & 1.73$\pm$0.05 &  85$\pm$2   &2.3\\
20-40\% & 0.45$\pm$0.01 &  152$\pm$1 & 2.00$\pm$0.05 &  69$\pm$2   &3.1\\
40-60\% & 0.41$\pm$0.01 &  151$\pm$1 & 2.39$\pm$0.09 &  51$\pm$1   &4.3\\
60-80\% & 0.35$\pm$0.01 &  150$\pm$1 & 3.19$\pm$0.12 &  35$\pm$1   &5.8\\
80-100\% & 0.26$\pm$0.01 &  147$\pm$1 & 4.90$\pm$0.20 &  19$\pm$1   &7.5\\
\end{tabular*}
\end{table}

\begin{table}[h]
  \caption{\label{tab:PbPb} Results for $\pi$, $K$, $p$ combined blast-wave fit with resonance decays for Pb--Pb \cmenn{2.76} data in momentum range $0.5<p_{ T} <3.0\,(\text{GeV}/c)$.}
\begin{tabular*}{\linewidth}{l@{\extracolsep{\fill}}c@{\extracolsep{\fill}}c@{\extracolsep{\fill}}c@{\extracolsep{\fill}}c@{\extracolsep{\fill}}c}
\hline
\hline
Centrality & $\left<\beta_T\right>$  & $T_\text{fo}$ (MeV) & $n$ & $dV/dy$ (fm${}^3$) & $\chi^2/\text{dof}$ \\
\hline
0-5\% & 0.66$\pm$0.01 &  149$\pm$1 & 0.34$\pm$0.04 &4273$\pm$122   &1.6\\
5-10\% & 0.65$\pm$0.01 &  149$\pm$1 & 0.39$\pm$0.04 &3585$\pm$100   &1.3\\
10-20\% & 0.64$\pm$0.01 &  149$\pm$1 & 0.44$\pm$0.04 &2682$\pm$74   &1.0\\
20-30\% & 0.62$\pm$0.01 &  150$\pm$1 & 0.55$\pm$0.05 &1796$\pm$51   &0.7\\
30-40\% & 0.59$\pm$0.01 &  150$\pm$1 & 0.71$\pm$0.06 &1188$\pm$35   &0.6\\
40-50\% & 0.55$\pm$0.01 &  151$\pm$1 & 0.96$\pm$0.07 & 756$\pm$23   &0.5\\
50-60\% & 0.50$\pm$0.01 &  152$\pm$1 & 1.26$\pm$0.09 & 428$\pm$14   &0.7\\
60-70\% & 0.45$\pm$0.01 &  151$\pm$1 & 1.74$\pm$0.12 & 235$\pm$9   &1.2\\
70-80\% & 0.39$\pm$0.02 &  150$\pm$1 & 2.30$\pm$0.17 & 117$\pm$5   &1.5\\
80-90\% & 0.34$\pm$0.01 &  148$\pm$1 & 3.08$\pm$0.12 &  55$\pm$2   &2.9\\
\end{tabular*}
\end{table}

\begin{table}[h]
  \caption{\label{tab:PbPb276TeV_MS} Results for $\pi$, $K$, $p$ combined blast-wave fit with resonance decays for Pb--Pb \cmenn{2.76} data in momentum range $0.5<p_{ T} <3.0\,(\text{GeV}/c)$.}
\begin{tabular*}{\linewidth}{l@{\extracolsep{\fill}}c@{\extracolsep{\fill}}c@{\extracolsep{\fill}}c@{\extracolsep{\fill}}c@{\extracolsep{\fill}}c}
\hline
\hline
Centrality & $\left<\beta_T\right>$  & $T_\text{fo}$ (MeV) & $n$ & $dV/dy$ (fm${}^3$) & $\chi^2/\text{dof}$ \\
\hline
0-10\% & 0.65$\pm$0.01 &  149$\pm$1 & 0.36$\pm$0.03 &3925$\pm$107   &1.4\\
10-20\% & 0.64$\pm$0.02 &  149$\pm$1 & 0.44$\pm$0.06 &2682$\pm$78   &1.0\\
20-40\% & 0.61$\pm$0.01 &  150$\pm$1 & 0.61$\pm$0.05 &1489$\pm$42   &0.6\\
40-60\% & 0.53$\pm$0.01 &  151$\pm$1 & 1.07$\pm$0.08 & 591$\pm$19   &0.6\\
60-80\% & 0.43$\pm$0.02 &  151$\pm$1 & 1.90$\pm$0.14 & 175$\pm$7   &1.3\\
\end{tabular*}
\end{table}

\begin{table}[h]
\caption{\label{tab:PbPb276TeV_MS_PCE150} Results for $\pi$, $K$, $p$ combined blast-wave fit with resonance decays in partial chemical equilibrium model with $T_\text{chem}=150\,\text{MeV}$ and $T_\text{kin}=T_\text{fo}$ for
Pb--Pb \cmenn{2.76} data in momentum range $0.5<p_{ T} <3.0\,(\text{GeV}/c)$..}
\begin{tabular*}{\linewidth}{l@{\extracolsep{\fill}}c@{\extracolsep{\fill}}c@{\extracolsep{\fill}}c@{\extracolsep{\fill}}c@{\extracolsep{\fill}}c}
\hline
\hline
Centrality & $\left<\beta_T\right>$  & $T_\text{fo}$ (MeV) & $n$ & $dV/dy$ (fm${}^3$) & $\chi^2/\text{dof}$ \\
\hline
0-10\% & 0.65$\pm$0.01 &  126$\pm$2 & 0.57$\pm$0.04 &5990$\pm$220   &0.7\\
10-20\% & 0.64$\pm$0.01 &  132$\pm$2 & 0.59$\pm$0.04 &3710$\pm$134   &0.6\\
20-40\% & 0.61$\pm$0.01 &  140$\pm$3 & 0.69$\pm$0.05 &1810$\pm$66   &0.5\\
40-60\% & 0.53$\pm$0.02 &  156$\pm$3 & 1.04$\pm$0.08 & 566$\pm$23   &0.6\\
60-80\% & 0.41$\pm$0.02 &  184$\pm$6 & 1.85$\pm$0.17 & 106$\pm$6   &0.9\\
\end{tabular*}
\end{table}

\begin{figure}
\centering
\includegraphics[width=\columnwidth]{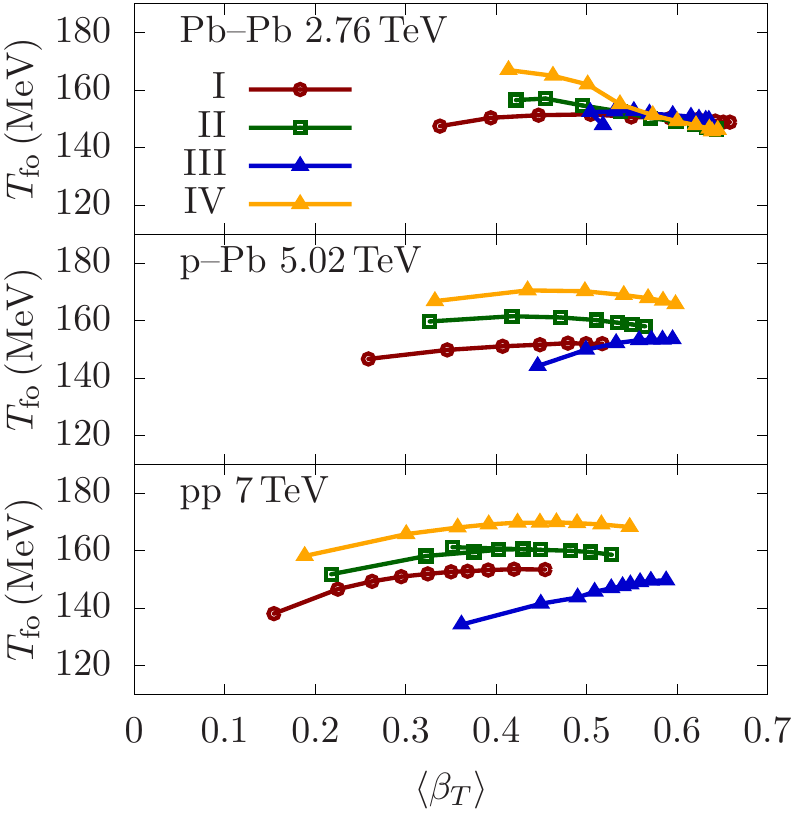}
\caption{Freeze-out temperature versus mean transverse flow for different
 $p_T$ fit ranges (see Table~\ref{tab:pT}) in a single freeze-out blast-wave model 
with resonance decays.}
\label{fig:plottkinvsbetatcombined}
\end{figure}

To study the model parameter sensitivity to different \pt regions, particle
spectra are fitted in different transverse momentum intervals as summarized in
Table~\ref{tab:pT}. In addition to our nominal range (I), we consider the
standard range (II), as well as high- and low-\pt regions (III, IV)  used in previous publications by ALICE~\cite{Abelev:2013vea}. We find that for a
given multiplicity, fit range (I) results in the lowest \mbeta in all cases except for
the most central Pb--Pb collisions as shown in
Fig.~\ref{fig:plottkinvsbetatcombined}. The freeze-out temperature measured in Pb--Pb collisions shows
little dependence on the fitting range, except at most peripheral bins. In smaller systems this dependence is
more pronounced and shows a decreasing trend as higher transverse momenta are
considered. 

{
\def\arraystretch{1.5}
\begin{table}
  \centering
  \caption{\label{tab:pT}Different choices of transverse momentum fit ranges (in GeV/$c$).}
  \begin{tabular*}{0.7\linewidth}{c|@{\extracolsep{\fill}}c@{\extracolsep{\fill}}@{\extracolsep{\fill}}c@{\extracolsep{\fill}}@{\extracolsep{\fill}}c}
   & $\pi$ & $K$ & $p$\\
   \hline
      I & [0.5,3.0] & [0.5, 3.0]& [0.5, 3.0]\\
     II & [0.5,1.0] & [0.2, 1.5]& [0.2, 3.0]\\
    III & [0.7,1.3] & [0.5, 1.5]& [1.0, 3.0]\\
     IV & [0.5,0.8] & [0.2, 1.0]& [0.3, 1.5]
  \end{tabular*}
\end{table}
}
\end{document}